\documentclass[aps]{revtex4}
\usepackage{graphicx}
\usepackage{amsmath}
\usepackage{amssymb}
\def\be{\begin{equation}}
\def\ee{\end{equation}}
\def\bea{\begin{eqnarray}}
\def\eea{\end{eqnarray}}
\def\pr{\partial}
\def\nno{\nonumber}
\def\bse{\begin{subequations}}
\def\ese{\end{subequations}}

\begin{document}
\title{The role of higher derivative bulk scalar in stabilizing a warped spacetime}

\author{Debaprasad Maity\footnote{E-mail: tpdm@mahendra@iacs.res.in} and Soumitra SenGupta\footnote{E-mail: tpssg@mahendra@iacs.res.in}}
\affiliation{Department of Theoretical Physics\\
Indian Association for the Cultivation of Science\\
Calcutta - 700 032, India
}
\author{Sourav Sur\footnote{E-mail: sourav.sur@uleth.ca}}
\affiliation{Department of Physics\\
University of Lethbridge\\
4401 University Drive, Lethbridge\\ 
Alberta - T1K 3M4, Canada
}
\begin{abstract}
The backreaction on the Randall-Sundrum warped spacetime is determined in presence of 
scalar field in the bulk. A general condition for the stability of such a model 
is derived for a bulk scalar field action with non-canonical higher derivative terms.  
It is further shown that the gauge hierarchy problem can be resolved in such a 
stabilized scenario by appropriate choice of various parameters of the
theory. The effective cosmological constant on the brane is shown to vanish.
\end{abstract}
\maketitle

\section{Introduction \label{intro}}
\noindent

Brane-world models have gained a considerable attention in recent years. A wide
number of applications of such models provide new ways to encounter some of the
unsolved problems in physics. From the particle physics point of view one of the
intriguing problems is the well-known `gauge hierarchy', which is related to the
mass renormalization of the standard model Higgs boson due to radiative corrections.
The Higgs boson appears naturally in the standard model (SM) so as to give
appropriate masses to the other fundamental bosons and fermions. While the mass
of the Higgs boson can not be stabilized due to the lack of symmetry, it is 
phenomenologically suggested to be in the range of TeV scale. The radiative 
corrections draw this TeV scale mass
to the Planck scale, leading to the well-known fine tuning problem of the Higgs 
mass. One possible solution of such a problem is supersymmetry, 
which is the symmetry between fermions and bosons. However, such a new symmetry 
leads to a large number of superpartner particles corresponding to all of the 
SM fundamental particles. Lack of experimental signatures of these 
superpartners till date, put serious constraints on supersymmetry. A possible 
alternative way of resolving the hierarchy problem is provided by the large or
warped compact extra dimensional models, proposed by Arkani-Hamed, Dimopoulus
and Dvali (ADD) and Randall-Sundrum (RS) \cite{arkani,lisa}. The main essence of 
such models is a geometric realization of the different scales in a theory. 
One important ingredient in this approach is the 
localization of all SM fields on a 3-dimensional hyper-surface, namely a 
`3-brane' \cite{arkani,lisa,witten,antoniadis,lykken,cohen,kaloper}, while only 
the fields coming from the gravity sector can propagate through the bulk. In the two-brane RS setup, 
the brane with negative tenson, i.e., the (visible) TeV brane on which all the SM fields live, is at 
one end while the brane with positive tenson, the so-called (hidden) Planck brane, 
is situated at the other end of the extra dimension. The geometry of the extra 
dimension is a circle modded out by the $Z_2$ symmetry, viz., $S^1/Z_2$ orbifold. 
The two ends of the extra dimension corresponds to the two orbifold fixed points.The model 
contains various parameters, the bulk cosmological constant $\Lambda$, the brane 
tensions $(V_{vis}, V_{hid})$ corresponding to the visible and hidden 
branes respectively and the brane separation $r_c$. Now, the static solution for 
the bulk Einstein equations \cite{lisa} leads to a relation between 
the bulk and brane cosmological constants $V_{hid} = - V_{vis} = \sqrt{- 24 \Lambda 
M^3}$ ($M$ being the 5-dimensional Planck mass) preserving the 4-dimensional 
Poincare invariance. For the desired warping one should have $k r_c \sim 12 ~(k = 
\sqrt{-\Lambda/(24 M^3)})$, where the bulk spacetime 
is an anti-de-sitter spacetime with a negative cosmological constant. 
If $k \sim$ Planck mass, then $r_c$ should have a  
value near the Planck length. To stabilize the value of this extra dimension, 
Goldberger and Wise (GW) \cite{gw} proposed a simple mechanism by introducing a  
massive scalar field with the usual (canonical) kinetic term in the bulk. 
The aim had been to obtain an effective four dimensional potential for the modulus $r_c$ 
by plugging back the 
classical solution for the scalar field in the bulk action and integrating out 
along the extra compact dimension. The minimum of this effective potential gives 
us the stabilized value of the compactification radius $r_c$. Several other 
works have been done in this direction \cite{luty-sundrum,maru,kogan,ferreira,csaki1,csaki2,ssg}.
However, in the calculation of GW \cite{gw}, the back-reaction of the 
scalar field on the background metric was ignored. Such back-reaction
was included later in ref. \cite{gub} and exact solutions for the background metric 
and the scalar field have been given for some specific class of potentials, motivated by 
five-dimensional gauged supergravity analysis \cite{freedman}.
Assigning appropriate values of the scalar field on the two branes, a stable value of the modulus $r$ was estimated
from the solution of the scalar field. 
It should however be noted  the scalar action in \cite{gub, csaki1} 
is inspired 
by a supergravity based model and has no correspondence with that in \cite{gw} even in a limiting sense.
These two models correspond to two different classes of bulk scalar actions such that for \cite{gub,csaki1} one can
calculate the exact warp factor with the full scalar back-reaction while \cite{gw} examines stability 
of the modulus by ignoring the back-reaction of the bulk scalar.\\
 
Another interesting work is in the context of a scalar field potential with a supersymmetric form, where 
it has been shown that the resulting model is stable\cite{smyth}\\

In a previous work \cite{addmssg}, starting from scalar bulk and boundary terms in the action, 
we re-examined the stability issue by generalizing the work of Goldberger and Wise \cite{gw}. 
Exploring into the region of the parameter space beyond the approximations adopted by Goldberger and Wise,  
we bring out some interesting features of the modulus potential and the corresponding stability 
conditions. 

In this work we examine the modulus stabilization by resorting to the usual modulus potential
calculation and it's subsequent minimization for a very general class of bulk scalar action. 
Keeping the non-canonical as well as higher derivative term in the bulk scalar action
we find the general condition for the modulus stabilization for   
back-reacted RS model \cite{lisa}. We exhibit the role of higher derivative terms 
in stabilizing the modulus as well as resolving the gauge hierarchy problem. 

The plan of this report is as follows: In section \ref{general} we begin with a
general bulk scalar action and find the condition on the back-reacted warp factor to achieve modulus stabilization
by estimating the derivative of the modulus potential without resorting to any specific choice for 
the bulk scalar.It is then followed by the solutions for the bulk scalar and the warp factor 
for specific choices for the bulk scalar action along the line of \cite{gub,csaki1,csaki2} 
in section \ref{gensol}. In section \ref{stab}, following the GW approach \cite{gw},
we perform a complete stability analysis of scalar back-reacted RS models \cite{lisa}
when the higher derivative and non-canonical terms are present. Here we determine the correlations among
the various parameters in the scalar potential 
along with the non-canonicity parameter and the corresponding stabilised value of the brane separation modulus $r$  
which resolves the electroweak gauge hierarchy problem.
Section \ref{cc} deals with the effective 4-dimensional 
cosmological constant for this kind of bulk action following refs \cite{sms,wald,rm,maeda-wands}. 
Finally, we make some concluding remarks in section \ref{conclu}.

\section{General Issues  \label{general}}

We start with a general action similar to that in our earlier work \cite{dmssgss}. In the 
last few years there have been many models where the presence of a bulk scalar field is
shown to have an
important role in the context of stability issue of brane-world scenario, bulk-brane
cosmological dynamics, higher dimensional black hole solutions and also in many other 
phenomenological issues in particle physics \cite{gw,gub,bulk1,bulk2,other,particle}.
Here we resort to a somewhat general type of self interacting scalar field along 
with the gravity in the bulk in order to analyze the stability of the RS type two-brane
model. We consider the following 5-dimensional bulk action 
\be \label{action}
S ~=~ \int d^5 x ~\sqrt{-g}\left[ - M^3 R ~+~ F(\phi, X) ~-~ V(\phi)\right] ~-~
\int d^4 x ~dy~ \sqrt{- g_a} \delta (y - y_a) \lambda_a (\phi) .
\ee
where $X = \pr_A {\phi} \pr^A {\phi}$, with `$A$' spanning the whole 5-dimensional bulk
spacetime. The index `$a$' runs over the brane locations and the corresponding brane 
potentials are denoted by $\lambda_a$. The scalar field is assumed to be only the function of extra 
spatial coordinate $y$.

Taking the line element in the form
\be \label{metans}
ds^2 ~=~ e^{- 2 A(y)} \eta_{\mu \nu} dx^\mu dx^\nu ~-~ dy^2 , 
\ee
where $\{y\}$ is the extra compact coordinate with radius $r_c$ such that 
$dy^2 = r_c^2 d\theta^2$, $\theta$ being the angular coordinate.
The field equations turn out to be
\bse \label{eq}
\bea
F_X \phi'' - 2 F_{XX} ~{\phi'}^2 \phi'' &=& 4 F_X \phi' A' - 
\frac {\pr F_X}{\pr \phi}{\phi'}^2 - \frac 1 2 \left(\frac {\pr F}
{\pr \phi} - \frac {\pr V}{\pr \phi}\right) + \frac 1 2 \sum_a \frac {\pr
\lambda_a(\phi)}{\pr \phi} \delta(y-y_a) \\
{A'}^2 &=& 4 C F_X ~{\phi'}^2 ~+~ 2 C ~\Big[F(X,\phi) ~-~ V(\phi)\Big] \\
A'' &=& 8 C F_X ~{\phi'}^2  ~+~ 4 C \sum_a \lambda_a(\phi) \delta (y - y_a) 
\eea
\ese
where 
\be
C ~=~ \frac 1 {24 M^3} ~;~ F_X ~=~ \frac {\pr F(X,\phi)}{\pr X} ~;~ 
F_{XX} ~=~ \frac {\pr^2 F(X,\phi)}{\pr X^2} \nno.
\ee
and prime $\{'\}$ denotes partial differentiation with respect to $y$. Two of the
above equations (\ref{eq}) are independent and the other one automatically follows 
from the energy conservation in the bulk.

The boundary conditions are
\bea \label{bcondition} 
2\left(F_X ~\phi' \right)|_{y=0} ~&=&~ \frac 1 2 
\frac{\pr \lambda_0(\phi_0)}{\pr \phi} ~~~;~~~ 
- 2\left(F_X ~\phi'\right)|_{y= \pi r_c} ~=~ \frac 1 2 
\frac{\pr \lambda_{\pi}(\phi_{\pi})}{\pr \phi} \\  
2 A'(y)|_{y =0} ~&=&~  4 C \lambda_0(\phi_0) ~~;~~
 -2 A'(y)|_{y = \pi r_c} ~=~  4 C \lambda_{\pi}(\phi_{\pi})
\eea
Now, without knowing the solutions of the above equations explicitly, we may 
analyze the stability of the modulus $r_c$, following the mechanism developed by Goldberger 
and Wise \cite{gw}. The brane separation $r_c$ in general is a dynamical variable 
associated with the metric component $g_{55}$. Integrating out the scalar field action 
over the extra coordinate $y$ in the background of the back-reacted five dimensional metric, 
the 4-dimensional effective potential for the brane separation
$r_c$ is obtained as
\bea
V_{eff} (r_c) &=& - 2 \int_0^{r_c\pi} dy ~ e^{- 4 A(y)} \Big[
- M^3 R ~+~ F(X,\phi) ~-~ V(\phi)\Big] 
 +~ e^{- 4 A(0)} \lambda_0 (\phi_0) ~+~ e^{- 4 A(r_c \pi)} 
\lambda_\pi (\phi_\pi) .
\eea

It may be noted that the effective potential is calculated with the warp factor $A(y)$, which takes care
of the full back reaction of the scalar field on the 5D metric through the equations of motion.
Therefore the effective potential for the modulus $r_c$ is 
calculated by integrating out the full action  
in the five dimensional background back-reacted metric. The modulus $r_c$ in general can be a dynamical variable and the
minimum of its effective potential determines the corresponding stable value. The role of bulk scalar field here is 
to stabilize the modulus associated with $g_{55}$ in 
the bulk five dimensional background spacetime. 

Now, using the above two boundary conditions and expression for the
potential from the above equation of motion 
\be
V(\phi) = - \frac 1 {2 C} {A'}^2 + 2 F_X ~{\phi'}^2 ~+~ ~F(X,\phi),
\ee
and also the expression for the Ricci scalar $R = 20 A'(y)^2 - 8 A''(y)$,
one gets the expression for the effective potential as
\bea
V_{eff} &=& - 16 M^3 \left[ A'(0) - A'(r_c \pi) e^{- 4 A(r_c \pi)} \right].
 \eea
 
Now, taking derivative with respect $y_{\pi} = \pi r_c$,
one finds, by the use of the equations of motion (\ref{eq}), 
the following algebraic
equation 

\be \label{minimum}
\frac {\pr V_{eff}(r_c)}{\pr (\pi r_c)} = 16 M^3 e^{-4 A(y)}
\left[A''(y) - 4 A'(y)^2\right]_{\pi r_c}
\ee
where $A (y)$ is specific solutions of Eqs.(\ref{eq}).

In order to have an extremum for $V_{eff}$ at some value of $r_c$, the right 
hand side of Eq.(\ref{minimum}) must vanish at that (stable) value of $r_c$.
This immediately implies that the value $A''(y)$ must be positive
and equals to the value of $4 A'(y)^2$ at $y = r_c$. Thus for different classes of 
back-reacted solutions for the warp factor for different bulk scalar actions, the above condition
determines the corresponding stable value of the modulus $r_c$.

\section{General Solution \label{gensol}}

Let us now resort to the general solutions of the full set of field equations (\ref{eq}).

We start with the case of a bulk scalar action with a simple non-canonical kinetic term, 
without any higher derivatives: 
\be \label{spmodel}
F(X,\phi) ~=~ f(\phi)~ X ,
\ee 
where $f(\phi)$ is any well-behaved explicit function of the scalar field $\phi$. Let us assume 
that $f(\phi) = \pr g(\phi)/\pr \phi$, where $g (\phi)$ is another explicit function of $\phi$. 
Then for a specific form of the potential
\be \label{nonpot}
V (\phi) ~=~ \frac 1 {16} \left(\frac {\pr g}{\pr \phi}\right)\left[ \frac{\pr W}{\pr g} \right]^2 ~-~ 
2 C ~W(g(\phi))^2
\ee
it is straightforward to verify, for some $W(g(\phi))$ and $g(\phi)$, that a solution to 
\be \label{soleq1}
\phi' ~=~ \frac 1 4 \frac{\pr W}{\pr g} ~;~~~
A' ~=~ 2 C ~W(g(\phi)) .
\ee
is also a solution to Eqs.(\ref{eq}), provided 
\be \label{bc}
\left[g(\phi)~\phi'\right]_a ~=~ \frac 1 2 \frac {\pr \lambda_a}
{\pr \phi_a}(\phi_a)~;~~~
\left[A'\right]_a ~=~ 4 C \lambda_a (\phi_a) .
\ee

Now, let us consider a more general case to include higher derivative term such as,
\be \label{genmodel}
F(X,\phi) ~=~  K(\phi) X ~+~ L(\phi) X^2 .
\ee
One of the motivations to consider this type of term in the Lagrangian originates from 
string theory \cite{mukhanov}. The low-energy effective string action contains higher-order 
derivative terms coming from $\alpha'$ and loop corrections, where $\alpha'$ is related to 
the string length scale $\lambda_s$ via the relation $\alpha' ~=~ \lambda_s/{2\pi}$. 
The 4-dimensional effective string action is generally given as
\be \label{stringaction}
S ~=~ \int d^4x ~ \sqrt{- \tilde g} \left[B_g(\phi) \tilde R ~+~ 
B_{\phi}^{(0)} X ~+~ \alpha' \left(c_1^{(1)} B_{\phi}^{(1)} X^2 ~+~ 
\cdots\right) ~~+~ {\cal O}\left(\alpha'^2\right)\right]
\ee
where $\phi$ is the dilaton field that controls the strength of the string coupling $g_s^2$ 
via the relation $g_s^2 = e^{\phi}$. In the weak coupling regime, the coupling function have 
the dependence $B_g \simeq B_{\phi}^{(0)} \simeq B_{\phi}^{(1)} \simeq e^{\phi}$. If we make 
a conformal transformation $g_{\mu\nu} = B_g(\phi) \tilde g_{\mu\nu}$, the string-frame action 
(\ref{stringaction}) is transformed to the Einstein-frame action \cite{mukhanov,gasperini} as :
\be
S_E ~=~ \int d^4x ~ \sqrt{-g} \left[\frac 1 2 R ~+~ K (\phi) X ~+~ 
L(\phi)X^2 ~+~ 
\cdots \right],
\ee
where
\be
K(\phi) ~=~ \frac 2 3 \left(\frac 1 {B_g} \frac{dB_g}{d\phi}\right)^2 ~-~ 
\frac {B_{\phi}^{(0)}}{B_g}, ~~~;~~~ L(\phi) ~=~ 2 c_1^{(1)} \alpha' B_{\phi}^{(1)}(\phi) .
\ee
Thus the 4-dimensional Lagrangian involves a non-canonical kinetic term for the scalar field, 
precisely in same form as that which we are considering here in 5-dimensions [Eq.(\ref{genmodel})]. 
With appropriate redefinition of the scalar field we can recast such a term in the Lagrangian as
\be \label{genF}
F(X, \phi) ~=~ f(\phi)[X ~-~ \beta X^2]
\ee
where $\beta$ is a constant parameter (in this case it is equal to unity) and $\{X, \phi\}$ are new 
variables in terms of old variables. This type of action is common in K-essence cosmological 
inflationary models \cite{kessence}. In what follows we will be using this K-essence type scalar action 
in the 5-dimensional bulk with a suitable potential function. The scalar field is, however, considered 
only a function of the extra (fifth) dimension $y$.

Now, assuming $f(\phi) = \pr g(\phi)/\pr \phi$, for a specific form of the potential
\be
V(\phi) ~=~ \frac 1 {16} \left(\frac {\pr g}{\pr \phi}\right)\left[ 
\frac {\pr W}{\pr h} \right]^2  + \frac {3 \beta}{256} \left(\frac 
{\pr g}{\pr \phi}\right) \left[\frac {\pr W}{\pr h}\right]^4- 2 C W^2
\ee
and for some arbitrary $W(h(g(\phi)))$ and $g(\phi)$, it is straightforward to 
verify that a solution to 
\be \label{soleq}
\phi' ~=~ \frac 1 4 \frac{\pr W}{\pr h} ~;~~~
A' ~=~ 2 C W(h(g(\phi))) ,
\ee
with the constraint relation
\be \label{relation}
\frac {dh} {dg} ~=~ {1 ~+~ \frac {\beta} 8 \left(\frac {\pr W}{\pr h}\right)^2} ,
\ee 
is also a solution to Eqs.(\ref{eq}), provided we have
\be \label{bc2}
\left[g(\phi)(1 ~- 2 \beta X)~\phi'  \right]_a 
~=~ \frac 1 2 \frac {\pr \lambda_a}{\pr \phi_a}(\phi_a)~;~~~
\left[A'\right]_a ~=~ 4 C \lambda_a (\phi_a) .
\ee
In the $\beta \rightarrow 0$ limit we at once get back the system of equations 
dealing with only the simple non-canonical kinetic term (\ref{spmodel}) in the 
scalar action, as discussed in the first part of this section. If in addition,
$f(\phi) \rightarrow 1/2$, one deals with the usual canonical kinetic term which 
has been discussed in detail in ref.\cite{gub}.

Considering now
\be \label{ansatz}
W(h) ~=~ \frac k {2 C} ~-~ u ~h^2(g(\phi)) ~~~;~~~~ g(\phi) ~=~ \alpha \phi,
\ee
where $k,~ u$ and $\alpha$ are the initial constant parameters of our model with 
their appropriate dimensions, Eq.(\ref{relation}) gives the solution for $h(g(\phi))$:
\be
h(g(\phi)) ~=~ \frac 1 {u \sqrt{\beta/2}} \tan \left(u \sqrt{\beta/2} ~\alpha \phi \right)
\ee
Clearly, in the limit $\beta \rightarrow 0$ we have $h = g$, which corresponds to what 
only the simple non-canonical kinetic term gives us [Eqs.(\ref{nonpot},~\ref{soleq1})]. 

From Eq.(\ref{soleq}) the solution for $\phi$ becomes
\be \label{solphi}
\sin \left(u \sqrt{\beta/2} ~\alpha \phi \right) ~=~ A_0 e^{- u \alpha y/2}
\ee
where $A_0 = \sin \left(u \sqrt{\beta/2} ~\alpha \phi_0 \right)$ 
with $\phi\vert_{y=0} \equiv \phi_0$. The solution for the warp factor $A(y)$ takes 
the form
\be \label{warp}
A (y) ~=~ ky ~-~ \frac {4C} {\beta u^2 \alpha} \ln 
\left({1 - A_0^2 e^{-u \alpha y}}\right) .
\ee
This is the exact form of the backreacted warp factor where the first term on the 
right hand side is same as that obtained by RS in absence of any scalar field and
the second term originates from the effect of backreaction due to the bulk scalar.

In the limit $\beta \rightarrow 0$ and $\alpha =$ const. 
\be
A (y) ~=~ k y ~+~ 2 C \alpha \phi_0^2 ~e^{-u \alpha y}
\ee
which is the exactly what has been discussed in ref.\cite{gub}.
It is now simple to determine the stable value for the modulus from using the minimization 
condition given by the Equ.(\ref{minimum}). 

\section{Stability Analysis \label{stab}}
Following the Goldberger-Wise mechanism \cite{gw}, we now analyze the stability of our
specific model, with $F (X, \phi)$ given by Eq.(\ref{genF})
in the preceding section. From Eq.(\ref{minimum}), we have the extremality condition
in a generic situation:
\be \label{mincond}
4 {A'}^2 ~- A'' ~=~ 0 .
\ee
This is an algebraic equation given in terms of the parameter $r_c$, i.e., the
compactification radius of extra dimension $y$. $A (y)$ is the
classical solutions of the field equations (\ref{eq}).
Now, putting the expressions for the various derivatives of
the metric solution $A$ in the above Eq.\ref{mincond}
one gets the following quadratic equation
\bea
{\cal Q} {\cal Y}^2 + {\cal P} {\cal Y} - \frac {\beta u^2 \alpha k^2}{C} = 0,
\eea
where, various notations are
\bea
{\cal Q} = u^2 \alpha^2 \left[ 1 - \frac {16 C}{\beta u^2 \alpha}\right]~~;~~ {\cal P} = u \alpha ( 8 k + u \alpha)~~;~~
{\cal Y} = \frac {A_0^2 e^{ - u \alpha y}}{1 - A_0^2 e^{ - u \alpha y}}
\eea
At this point it is worth noticing that numerical value of ${\cal Y}$
should be positive.

From the above equation clearly, we can have two different cases.

{\bf Case i}) $ {\cal Q} > 0$, then
the above equation has only one real solution considering the
fact that ${\cal Y} > 0$. The corresponding root is
\bea
 {\cal Y} = \frac 1 {2 {\cal Q}} \left( \sqrt{{\cal P}^2 + \frac {4 {\cal Q} \beta u^2 \alpha k^2}{C}} - {\cal P}
 \right)
 \eea
which gives us the maximum of the potential. So, the point we get
is unstable.

{\bf Case ii})  $ {\cal Q} < 0$, which in turn
says $ \beta << 1$ provided $u$ $\sim$ Plank scale,
then we have two solutions
\bea
{\tilde{\cal Y}}_{\pm} = \frac 1 {2 |{\cal Q}|} \left(  {\cal P} \pm \sqrt{{\cal P}^2 -
\frac {4 |{\cal Q}| \beta u^2 \alpha k^2}{C}}
\right)
\eea
where $|{\cal Q}|$ is the absolute value of ${\cal Q}$. As we have
checked that the larger value of ${\cal Y}  = {\tilde{\cal Y}}_+$ gives
us the stable point for the modulus. So, naturally, the lower
value ${\cal Y}  = {\tilde{\cal Y}}_-$ gives the unstable point.

So, corresponding to these minimum ${\tilde{\cal Y}}_+$ and maximum
${\tilde{\cal Y}}_-$ of the effective radion potential, one
gets respective distance moduli $y_{\pi}^{\pm}$ as
\bea
y_{\pi}^{\pm}  = \frac 1 {u \alpha} log\left[\frac
{\left({\cal P} + 2 |{\cal Q}| \pm \sqrt{{\cal P}^2 -
\frac {4 |{\cal Q}| \beta u^2 \alpha k^2}{C}} \right) A_0^2}
{{\cal P} \pm \sqrt{{\cal P}^2 - \frac {4 |{\cal Q}| \beta u^2
\alpha k^2}{C}} } \right]
\eea

Now, we are interested to study our solution for the metric
as well as scalar field at this stable point. Before we
start, it is useful to define some dimensionless parameters out
of the various know dimensionful parameters as
\bea
m = k/(u \alpha)~~~;~~~~n = {4 C}/(u^2 \beta \alpha),
\eea
From now on, we will read out the various expressions in term of these
new parameters. We are interested to study the metric
at the stable point $y_{\pi}^+ = k \pi r_s$, where $r_s$ is the said
to be stable distance between the two branes.
\bea
{\tilde {\cal Y}}_+ &=& \frac 1 {2(4n - 1)}\left[ (1 + 8 m) +
\sqrt{(1 + 8 m)^2 - 16(4n - 1) \frac {m^2} {n}} \right] \\
y_{\pi}^{+} &=& \frac m k~\ln\left[\frac {(1 + {\tilde {\cal Y}}_+) A_0^2}
{{\tilde {\cal Y}}_+}\right]
\eea

So, by using the above expressions, we get the expression for stable
value of $A(r_s)$
as

\bea
A(r_s) = m~ \ln\left[\frac {(1 + {\tilde {\cal Y}}_+) A_0^2}
{{\tilde {\cal Y}}_+}\right] -
n~\ln\left[\frac 1 {1 +{\tilde {\cal Y}}_+}\right]
\eea
\begin{figure}
\hfill
\begin{minipage}{.48\textwidth}
\includegraphics[width=3.20in,height=3.0in]{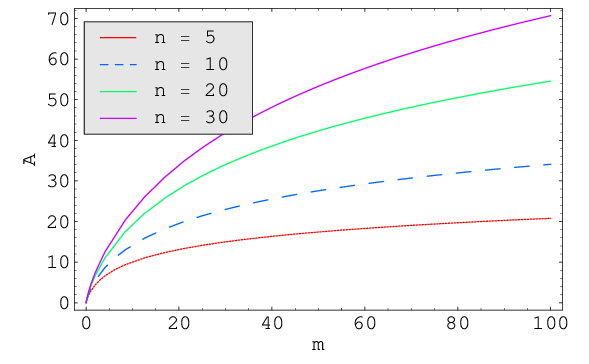}
\caption{Variation of
warp factor of the metric $A(r_s)$ with parameter
$m = k/(u \alpha)$ setting fixed values
$\{= 5, 10, 20 ~\hbox{and} ~30 \}$ of the other parameter
$ n = {4 C}/(u^2 \beta \alpha)$ and $A_0 = \sin \left(u {\sqrt{
{\beta}/{2}}} \alpha \phi_0 \right) = 1$.} \label{fig1}
\end{minipage}
\hfill
\begin{minipage}{.48\textwidth}
\includegraphics[width=3.20in,height=3.0in]{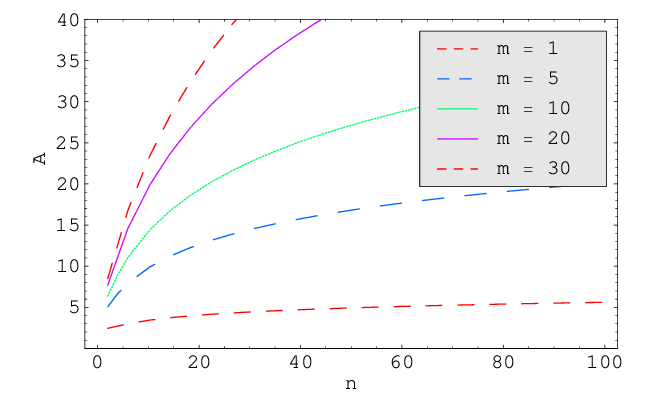}
\caption{Variation of warp factor of the metric $A(r_s)$ with parameter
$n = {4 C}/(u^2 \beta \alpha)$ setting fixed values $\{ = 1, 5, 10, 20
~\hbox{and}~  30 \}$ of the other parameter $m = k/(u \alpha)$ and
$A_0 = \sin \left(u~{\sqrt{ {\beta}/{2}}} ~\alpha \phi_0 \right) = 1$.} \label{fig2}
\end{minipage}
\hfill
\end{figure}
\begin{figure}
\hfill
\begin{minipage}{.48\textwidth}
\includegraphics[width=3.20in,height=3.0in]{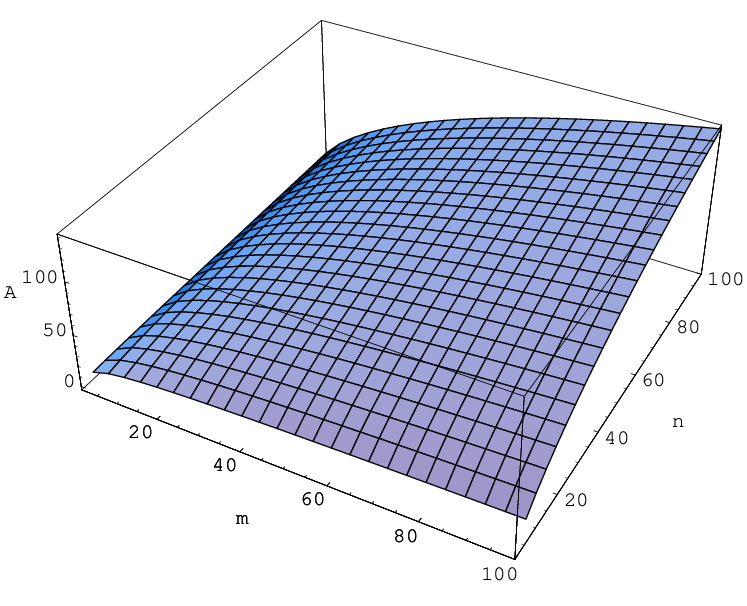}
\caption{Variation of warp factor of the metric $A(r_s)$ with both
the parameters $m = k/(u \alpha)$ and $n = {4 C}/(u^2 \beta \alpha)$
with $A_0 = \sin \left(u {\sqrt{{\beta}/{2}}} \alpha \phi_0 \right) = 1$.}
\label{surface}
\end{minipage}
\hfill
\begin{minipage}{.48\textwidth}
\includegraphics[width=3.20in,height=3.0in]{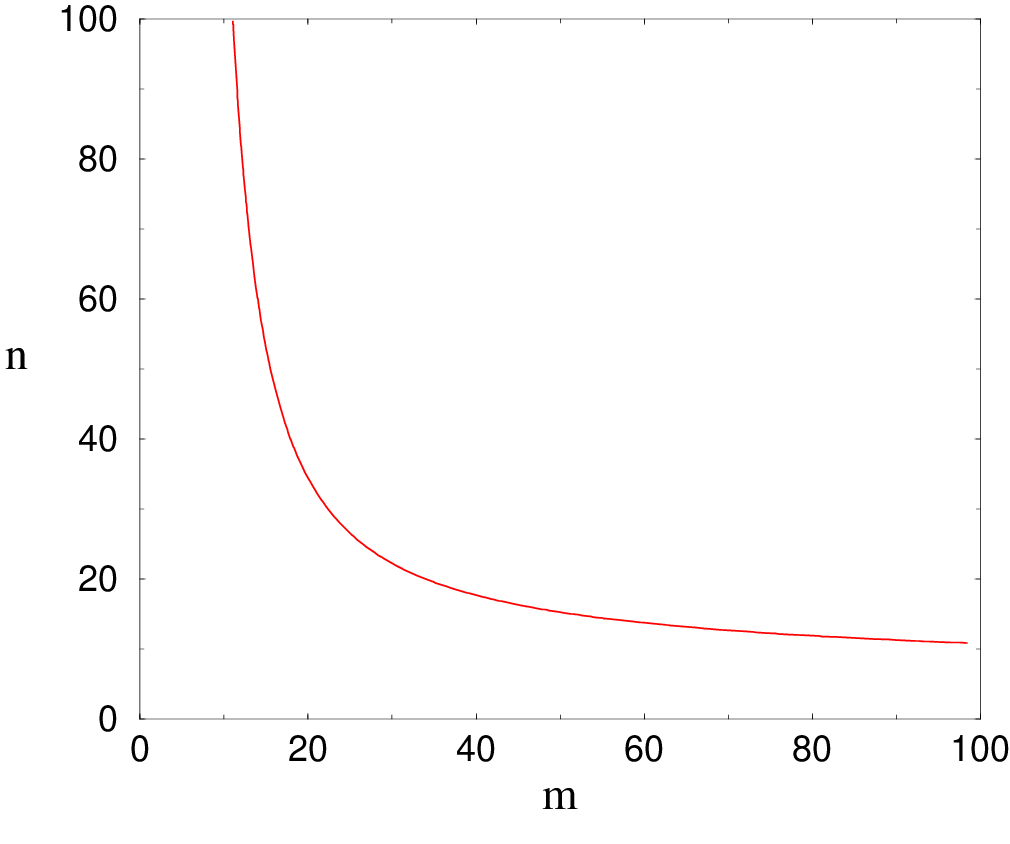}
\caption{Variation of $n = {4 C}/(u^2 \beta \alpha)$ with
the variation of $m = k/(u \alpha)$ keeping fixed the value of
warp factor $A(r_s) = 36$ which approximately gives the scaling
down of Planck scale to TeV scale.} \label{fig4}
\end{minipage}
\hfill
\end{figure}
\begin{figure}
\includegraphics[width=3.20in,height=3.0in]{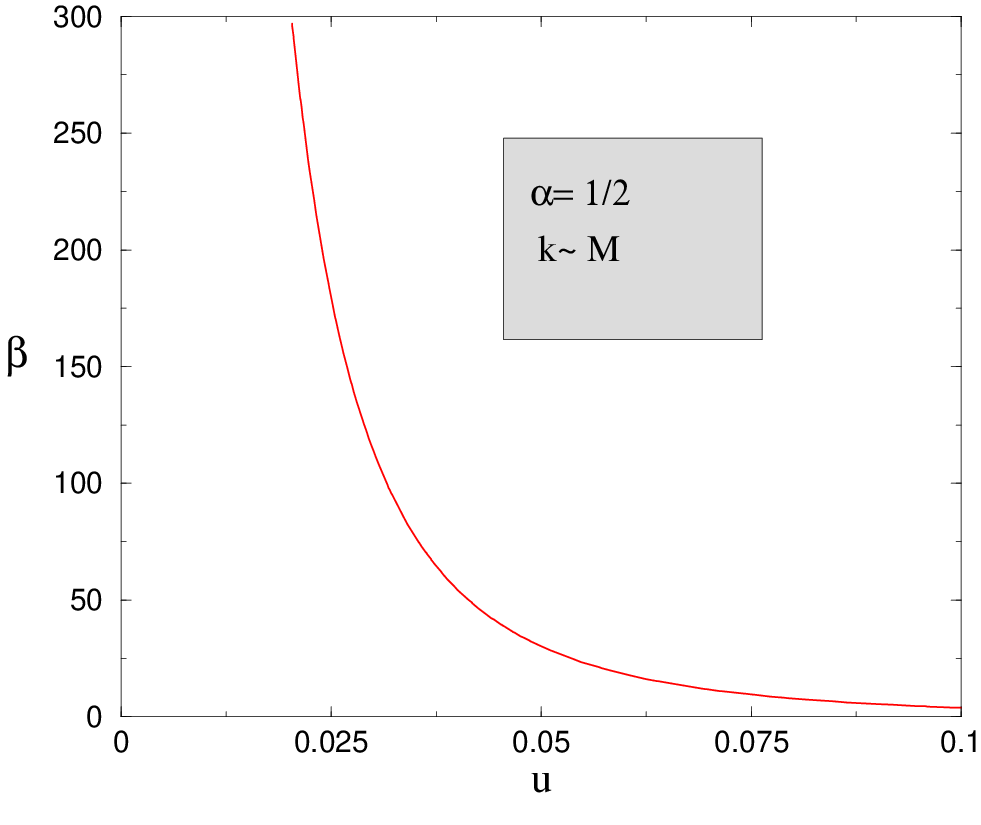}
\caption{Variation of $\beta$ with $u$ for which we get $A(r_s)
\sim 36$ with fixed values
 for $\alpha = 1/2 ~\hbox{and}~ k \sim M $.} \label{ubeta}
 \end{figure}

 So, from the above expression for the stable value of warp factor
 $A(r_s)$, we can calculate the values of the set of parameters
 $\alpha, \beta, u, k  ~\hbox{and}~ \phi_0$ of our model from various
 phenomenological constraints. For simplicity we will do
 our analysis by setting the parameter $A_0 = \sin \left(u~{\sqrt{
 {\beta}/{2}}} \alpha \phi_0 \right) = 1$ .

 The main issue addressed in the RS two-brane model \cite{lisa} is the
 large mass hierarchy from an extra dimension whose compactification
 radius is as small as of the order of Planck length ($l_{p} \sim
 10^{-33}$ cm).In the present scenario, because of the scalar field in the bulk, the
 the modified warp factor
 $A(y)$ is given by Eq.(\ref{warp}). Now, to get the acceptable hierarchy
 between the two fundamental scales, the modulus $r_c$ must be stabilized to a value $r_s$
 such that the warp factor $A(r_s) \sim 36$. Figs. \ref{fig1} and
 \ref{fig2} depicts the variation of the value of $A(y)$ with the parameter $m$ ( for fixed $n$) and $n$ ( for fixed $m$)

whereas in Fig. \ref{surface} we show the surface plot of
the warp factor versus the two dimensionless variables $m = k/(u\alpha)$
and $n = 4C/(u^2\beta\alpha)$. Finally, a numerical plot between
$m$ and $n$ corresponding to a fixed value $A(r_s) = 36$ up to an
error $\sim 2 \times 10^{-8}$ is shown in Fig. \ref{fig4}. It is evident
that we can get an infinite number of values for the above parameters
by which we can resolve the hierarchy problem in connection with the
Higgs mass.

Now, we will try to estimate the values of the parameters $\alpha, \beta,
u, k ~\hbox{and}~ \phi_0$ for some values of $m$ and $n$. Since, we have
a total of five free parameters in our model, we can specify any two of
them by suitably defining a combination of all of them. Let us choose $k$
and $\alpha$ from the physical point of view, to be $\alpha = 1/2$ (making
the scalar kinetic term to be canonical) and $k \sim M$. Then, we get the
plot shown in Fig. \ref{ubeta}. It is
interesting to note that for values of $\beta \ll \alpha$ the stability
requirement of the brane-world model is satisfied.

\section{Effective Cosmological Constant \label{cc}}
\noindent

A particularly elegant way of calculating the effective induced brane 
world gravity was proposed by Shiromizu, Maeda and Sasaki\cite{sms}. 
The method contains the projections of the bulk equation of 
motions for the different bulk fields  
on the brane sitting at the orbifold fixed points along the 
extra coordinate. 
The basic idea is to use the Gauss-Codacci equation to project the 
5D curvature, of any general metric, along the embedded brane. This is 
a covariant approach to get the brane equation of motion(general
formalism is given in reference\cite{wald} in details).
Here briefly describing the formalism 
following \cite{sms,rm,langlois} we apply this to our specific model discussed 
in the previous sections.

In the brane world scenario, we describe our 4-dimensional world by a 
3-brane $(V, h_{\mu\nu})$ in 5-dimensional spacetime $(M, g_{\mu\nu})$. 
Now, if we take $n_{\mu}$ is the unit normal to the hyper-surface $V$ 
along the Gaussian normal coordinate $y$ such that 
$n_{\mu} d{\cal X}^{\mu} ~=~ y$, then the induced metric on the 
3-brane $h_{\mu\nu}$ will be
\be
h_{\mu\nu} ~=~ g_{\mu\nu} + n_{\mu} n_{\nu} .
\ee
with the line element for the bulk, $ds^2 ~=~g_{\mu\nu} d{\cal X}^{\mu}
d{\cal X}^{\nu}$.

\noindent
Different 
projections of the bulk curvature tensor, follows from Gauss equation, 
lead to the following 4D curvatures 
\bse \label{projection}
\bea
R^{\gamma}_{~\delta\sigma\rho} ~&=&~ {\cal R}^{\alpha}_{~\beta\mu\nu} 
h_{\alpha}^{~\gamma} h_{\delta}^{~\beta} h_{\sigma}^{~\mu}
h_{\rho}^{~\nu} ~-~ K_{\sigma}^{~\gamma} K_{\rho\delta}
~+~ K_{\nu}^{~\beta} K_{\sigma\delta} \\
R_{\sigma\rho} ~&=&~ {\cal R}_{\mu\nu} h_{\sigma}^{\mu} h_{\rho}^{\nu} 
~+~ {\cal R}^{\alpha}_{~\beta\mu\nu} n_{\alpha} h_{\sigma}^{~\beta} n^{\mu}
h_{\rho}^{~\nu} ~-~ K K_{\sigma\rho} ~+~ K_{\rho}^{~\mu} K_{\mu\sigma} \\
R ~&=&~ {\cal R} ~+~ 2 {\cal R}_{\mu\nu} n^{\mu}n^{\nu} ~-~ K^2 
+ K^{\mu\nu}K_{\mu\nu} 
\eea
\ese
where $\cal R$ and  $R$ are curvature scalar of 5D manifold 
$(M,g_{\mu\nu})$ and 3-brane 
namely domain wall $(V,h_{\mu\nu})$ respectively. The extrinsic curvature 
of $V$ is denoted by
\be
K_{\mu\nu} ~=~ h_{\mu}^{~\alpha} h_{\nu}^{~\beta} \nabla_{\alpha} 
n_{\beta} ~~~~;~~~
K ~=~ K_{\mu}^{~\mu}
\ee
where $\nabla_{\mu}$ is the covariant differentiation with respect to 
$g_{\mu\nu}$ and the Gauss-Codacci equation determines the change of 
extrinsic curvature along 
$y$ axis via
\be \label{gauss-codazzi}
D_{\mu} K_{\nu}^{~\mu} ~-~ D_{\mu} K ~=~ {\cal R}_{\alpha\beta} 
n^{\alpha} h_{\nu}^{~\beta}
\ee
where $D_{\mu}$ is the covariant differentiation with respect to $h_{\mu\nu}$. 

Now, using all the projections of the bulk curvatures Eq.\ref{projection}, 
the 5D Einstein equation is given by, 
\be
{\cal R}_{\mu\nu} - \frac 1 2 g_{\mu\nu} {\cal R} ~=~ \kappa^2 
{\cal T}_{\mu\nu} ,
\ee
where ${\cal T}_{\mu\nu}$ is the bulk energy-momentum tensor and 
$\kappa^2 ~=~ 1/M^3$. The sum of various decompositions 
of the bulk Riemann tensor 
\be
{\cal R}_{\mu\alpha\nu\beta} ~=~ \frac 2 3 \left( g_{\mu[\nu} 
{\cal R}_{\beta]\alpha} ~-~  g_{\alpha[\nu} {\cal R}_{\beta]\mu} \right)
 ~-~ \frac 1 6 g_{\mu[\nu} g_{\beta]\alpha}
{\cal R} ~+~ {\cal C}_{\mu\alpha\nu\beta},
\ee
where ${\cal C}_{\mu\alpha\nu\beta}$ is known as Wyel tensor which is 
the traceless part of the Riemann tensor. All these lead to the Einstein's equation on the 3-brane $V$ as
\be \label{inducedeq}
G_{\mu\nu} ~=~ \frac 2 {3 M^3} \left[ {\cal T}_{\alpha\beta} ~-~ 
\left ( {\cal T}_{\gamma\delta} n^{\gamma} n^{\delta} ~+~ \frac 1 4 
{\cal T} \right ) 
g_{\alpha\beta} \right] h_{\mu}^{~\alpha} h_{\nu}^{~\beta} ~+~ 
{\cal E}_{\mu\nu} ~+~ F_{\mu\nu} 
\ee
where
\be
{\cal E}_{\mu\nu} = {\cal C}^{\sigma}_{~\alpha\beta\gamma} 
n_{\sigma} n^{\beta} h_{\mu}^{~\alpha} h_{\nu}^{~\gamma}  ~~;~~
F_{\mu\nu} = - k k_{\mu\nu} + K_{\mu}^{~\rho} K_{\nu\rho} + 
\frac 1 2 (K^2 - k_{\alpha\beta} K^{\alpha\beta} ) h_{\mu\nu} \nno
\ee
>From these the Gauss-Codacci equation turns out to be
\be \label{gc}
D_{\mu} K_{\nu}^{~\mu} ~-~ D_{\mu} K ~=~ \kappa^2 {\cal T}_{\alpha\beta} 
n^{\alpha} h_{\nu}^{~\beta}
\ee
Following the work in the context of a dilaton coupled gravity theory  \cite{maeda-wands}, 
we now use this formalism in our scenario where we have higher derivative terms for the scalar field.
>From the action Eq.\ref{action} we have the bulk energy-momentum tensor
is
\be
{\cal T}_{\mu\nu} ~=~ F_X \pr_{\mu} \phi \pr_{\nu} \phi ~-~ 
\frac 1 2 [F(X,\phi) 
~-~ V(\phi)] g_{\mu\nu}
\ee
where bulk scalar field $\phi$ depends only on the extra coordinate $y$.

In order to explore the induced Einstein equation of motion, 
we first need to calculate the extrinsic curvature $K_{\mu\nu}$ of 
the brane by using the junction condition. If, the brane, on which we 
are interested in, has the energy momentum tensor $T_{\mu\nu}$ 
then from Israel junction condition\cite{israel} at the location 
( say at $ y ~=~ y_i$) of the brane is 
\be
[K_{\mu\nu}]_i ~=~ - \kappa^2 \left( T_{\mu\nu} ~-~ 
\frac 1 3 h_{\mu\nu} T \right)_i
\ee
Now, in general we can split the brane energy-momentum tensor to be
\be                                                     
T_{\mu\nu} ~=~  \frac 1 2 \lambda h_{\mu\nu}  + \tau_{\mu\nu} 
\ee
where $\lambda$ is brane tenson and $ \tau_{\mu\nu} $ is the energy-momentum 
tensor coming from localized matter Lagrangian on the brane.
Using all these, we get the corresponding extrinsic curvature tensor 
and scalar as,
\be \label{extrinsic}
K_{\mu\nu} = - \frac {\kappa^2}{2} \left[ - \frac 1 6 h_{\mu\nu} \lambda 
+ \left (\tau_{\mu\nu} - \frac 1 3 h_{\mu\nu} \tau \right) \right]~~;~~
K = \frac {\kappa^2}{6} ( 2 \lambda + \tau ) 
\ee                                              
where $\tau ~=~ \tau_{\mu}^{~\mu}$, trace of the energy-momentum tensor.
So, the expression for $F_{\mu\nu}$ will be
\be
F_{\mu\nu} ~=~ \frac {\kappa^4}{48} \lambda^2 ~+~ \frac {\kappa^4}{12} 
\lambda \tau_{\mu\nu}
~+~ \kappa^4 \pi_{\mu\nu}
\ee
where 
\be \label{higherorder}
\pi_{\mu\nu} = - \frac 1 {12} \tau \tau_{\mu\nu} + \frac 1 4 
\tau_{\mu}^{~\alpha} 
\tau_{\nu\alpha} - \frac 1 {24} h_{\mu\nu} \left( 3 \tau_{\alpha\beta}
\tau^{\alpha\beta}
-  \tau^2 \right) \nno
\ee
Finally, we arrive at the induced Einstein equation of motion on the 
3-brane substituting the equations Eq.\ref{extrinsic}, Eq.\ref{higherorder} 
on the Eq.\ref{inducedeq}
\be \label{induced}
 G_{\mu\nu} = \frac 2 {3 M^3} \left[ {\cal T}_{\alpha\beta} - 
\left ( {\cal T}_{\gamma\delta} n^{\gamma} n^{\delta} + 
\frac 1 4 {\cal T} \right ) g_{\alpha\beta} \right] h_{\mu}^{~\alpha} 
h_{\nu}^{~\beta} + {\cal E}_{\mu\nu} + 
\frac {\kappa^4}{48} \lambda^2 + \frac {\kappa^4}{12} \lambda \tau_{\mu\nu}
+ \kappa^4 \pi_{\mu\nu} 
\ee
Taking brane matter energy momentum tensor $\tau_{\mu\nu}$ and hence  $\pi_{\mu\nu}$ to be zero, the
contribution from the four dimensional energy-momentum
tensor comes only from the brane tension and is given by,
\be
T_{\mu\nu} ~=~ \frac 1 2 \lambda(\phi)~h_{\mu\nu}.
\ee
Where $\lambda(\phi)$ is the brane tenson which in our 3-brane model is negative. 

The expression for $F_{\mu\nu}$, which is quadratic in extrinsic 
curvature $(K_{\mu\nu} ,K)$ is
\be \label{F}
F_{\mu\nu} ~=~ \frac 1 {48 M^6} \lambda(\phi)^2 h_{\mu\nu} .
\ee

Since, the stabilizing scalar field $\phi$ depends only on the  
extra coordinate $y$, the bulk energy-momentum tensor ${\cal T}_{\mu\nu}$ 
contributes to the induced Einstein equation Eq.\ref{induced} in the following way :
\be 
{\cal T}_{\mu\nu} n^{\mu} n^{\nu} + \frac 1 4 {\cal T} = 
F_X \pr_{\mu} \phi \pr_{\nu} \phi \left(n^{\mu} n^{\nu} + \frac 1 4 g_{\mu\nu} \right) 
- \frac 1 2 \left(F(X,\phi) - V(\phi)\right) \left(g_{\mu\nu} 
n^{\mu} n^{\nu} + \frac 1 4 \right) .
\ee
This readily gives, by using the relation 
$g_{\mu\nu} = h_{\mu\nu} - n_{\mu} n_{\nu}$ and the property
 $h_{\mu}^{~\nu} n_{\nu} ~=~ 0$,
\be \label{T1}
{\cal T}_{\mu\nu} n^{\mu} n^{\nu} + \frac 1 4 {\cal T} ~=~ 
\frac 3 4 F_X \phi'^2 ~-~ 
\frac 1 8 (F(X,\phi) ~-~ V(\phi)) .
\ee
and
\be \label{T2}
{\cal T}_{\alpha\beta} h_{\mu}^{~\alpha} h_{\nu}^{~\beta}  ~=~ - \frac 1 2  
(F(X,\phi) ~-~ V(\phi)) h_{\mu\nu} .
\ee

Now, by putting Eq.(\ref{F}, \ref{T1}, \ref{T2}) and 
$\pi_{\mu\nu} ~=~ \tau_{\mu\nu} ~=~ 0$ in
the effective induced Einstein equation Eq.\ref{induced}, one obtains
\be \label{finalinduced}
G_{\mu\nu} ~=~ \frac 1 {M^3} \left[ \frac 1 {48 M^3} \lambda(\phi)^2 ~-~ 
\frac 1 4 (F(X,\phi) ~-~ V(\phi)) ~-~ F_X \phi'^2 \right] h_{\mu\nu} 
~+~ {\cal E}_{\mu\nu} .
\ee
The term appearing as the coefficient of  $h_{\mu\nu}$ plays the role of effective 
four dimensional cosmological constant (say $\Lambda$). It is an
interesting point to note that so far no explicit expression for $F(X,\phi)$ has been assumed.

Eq.\ref{finalinduced} reduces to
\be
 G_{\mu\nu} ~=~ - \Lambda h_{\mu\nu} ~+~ {\cal E}_{\mu\nu} .
\ee
where,
\be 
\Lambda ~=~ - \frac 1 {48 M^3} \lambda(\phi)^2 ~+~ \frac 1 4 (F(X,\phi) 
~-~ V(\phi)) ~+~ F_X \phi'^2 
\ee
Applying equation of motions Eq.\ref{eq} and the corresponding 
boundary conditions
Eq.\ref{bcondition} the effective cosmological constant becomes zero. 
so that the final equation of motion reduces to
\be
G_{\mu\nu} ~=~ {\cal E}_{\mu\nu} .
\ee
The key modification to the induced Einstein equation is the bulk 
effect when there is no localized energy momentum tensor on the brane. 
The Weyl curvature
term ${\cal E}_{\mu\nu}$ induces the correction from 5D gravity wave effects. This  
term can not be calculated by the brane data only and we have to solve the 
full bulk equation of motion for the metric and calculate the 
corresponding KK modes.

\section{Conclusion   \label{conclu}} 
Starting from a general action with higher derivative, non-canonical kinetic
term for a bulk scalar in a RS-type two brane model, we have derived the modulus stabilization condition on the
scalar action in a back-reacted warped geometry. This result generalizes our
earlier work\cite{dmssgss} where a specific form of the higher derivative term namely tachyon-like action
was shown to stabilize the model. Calculating the exact form of the
warp factor with full back-reaction, the naturalness issue in the context of Higgs mass has been explored. 
A large set of values of the parameters of the bulk scalar potential and the parameters responsible for 
the presence of non-canonical and 
higher derivative terms are shown to produce the desired warping from the Planck scale to Tev scale
as a solution to the well known gauge hierarchy problem. The correlation among these parameters are 
determined. 
The induced cosmological constant on the 3-brane is shown to vanish in such a model.
This work thus determines the general conditions on a bulk scalar action in a back-reacted Randall-Sundrum brane-world 
so that the problems of 
stabilizing the modulus ,gauge hierarchy as well as a  vanishing effective cosmological 
cosmological constant can all be resolved together.

\acknowledgments
DM acknowledges Council of Scientific and Industrial Research, Govt. of India for
providing financial support. 
 

\end{document}